\documentclass[%
 aip,
 jmp,%
 amsmath,amssymb,
 reprint,%
]{revtex4-2}

\usepackage{graphicx}
\usepackage{dcolumn}
\usepackage{bm}

\begin{document}

\preprint{AIP/123-QED}

\title{ An epsilon-near-zero-based Dallenbach absorber }

\author{Viacheslav V. Medvedev}

 \altaffiliation[Also at ]{Moscow Institute of Physics and Technology (State University), Institutskii pereulok 9, Dolgoprudnyi 141701, Moscow region, Russia}
 \email{medvedev@phystech.edu}
 \affiliation{ 
Institute of Spectroscopy of the Russian Academy of Science, Fizicheskaya 5, Troitsk, Moscow 108840, Russia}%

\date{\today}

\begin{abstract}
We report a theoretical analysis of total absorption conditions in a structure consisting of a lossy coating layer on top of a specular metal substrate, which is known as a Dallenbach absorber. All possible combinations of complex refractive indices of the coating material and thicknesses of the coating layer providing complete absorption are described for the case of normal incidence of a plane wave. The work of coatings based on epsilon-near-zero (ENZ) materials is analyzed in detail. Simple analytical relations are obtained for the conditions of total absorption. It is shown that the range of angles where strong absorption occurs is limited by the phenomenon of total external reflection. The characteristics of coatings based on indium-tin oxide, which is an example of an ENZ material, are analyzed. 
\end{abstract}

\maketitle

\section{Introduction}

\begin{figure}[htbp]
\centering
\includegraphics[width=0.5\linewidth]{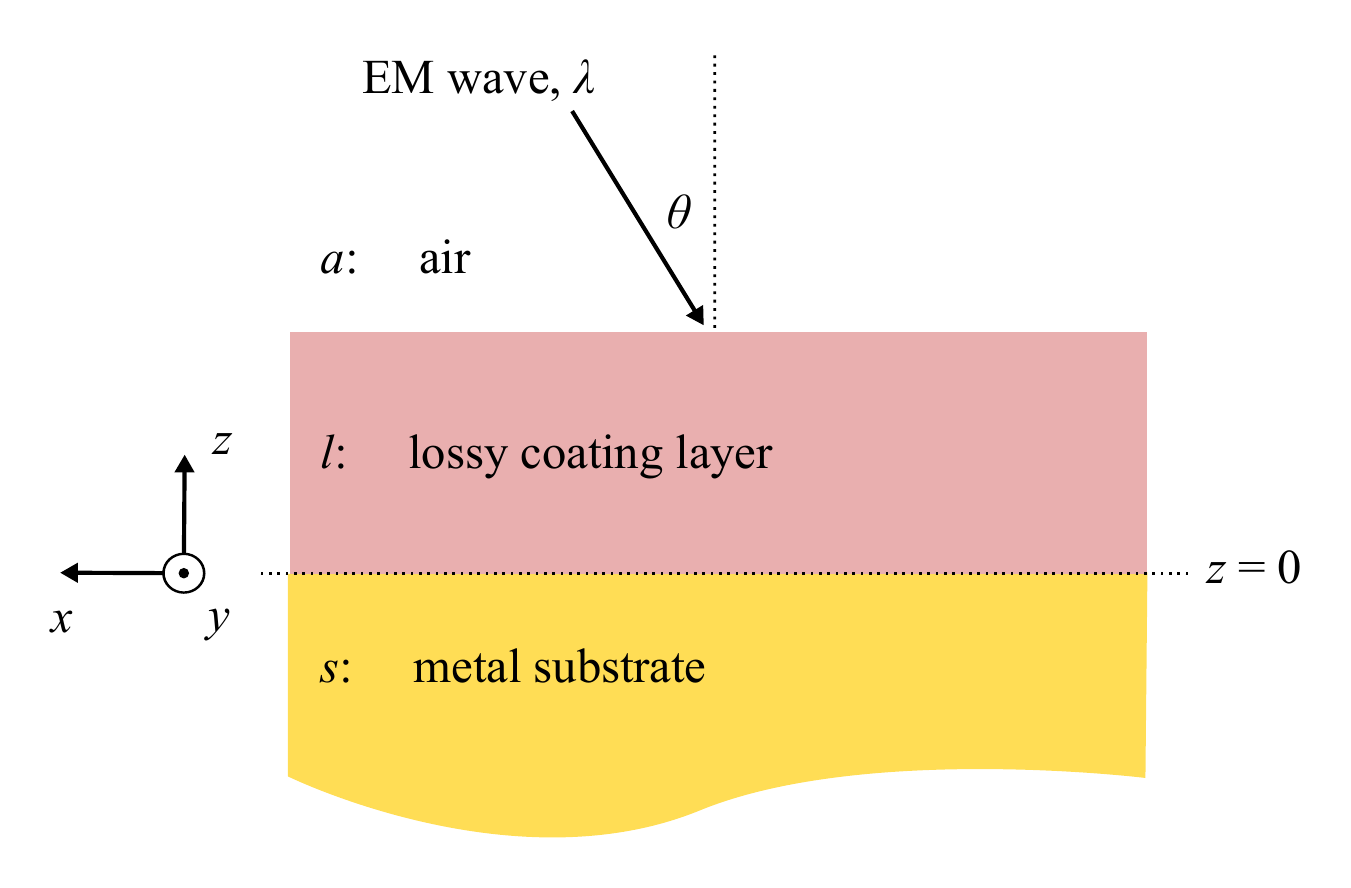}
\caption{Schematic diagram of the Dallenbach absorber.}
\label{fig_1}
\end{figure}

A Dallenbach absorber is one of the most striking and simplest devices designed to effectively absorb electromagnetic radiation \cite{Dallenbach1938, Knott2004,Saville2005Review,Tretyakov2016IEEE}. According to the original design, it consists of a homogeneous lossy coating layer placed on top of a reflecting metal substrate (see Fig. \ref{fig_1}). Dallenbach absorbers benefit from the destructive interference that minimize reflection and, as a result, maximize absorption of radiation incident on a structure. For a normally incident monochromatic plane wave with a wavelength $\lambda$, one can write approximate equations for unity absorption conditions that relate the optimal refractive index ($n_l$), extinction coefficient ($\kappa_l$) and thickness ($d_l$) of the coating layer:
\begin{subequations}
\label{eq:Dbch}
\begin{eqnarray}
d_l = \frac{\lambda(2m+1)}{4n_l},
\label{subeq:2}
\end{eqnarray}
\begin{equation}
\kappa_l = \frac{2}{\pi(2m+1)},
\label{subeq:1}
\end{equation}
\label{eq_1}
\end{subequations}
where $m=0,1,2,...$. These equations are derived for an idealized case when the metal substrate is approximated by a perfect electric conductor (PEC). The integer $m$ in Eq. \ref{eq_1} is referred to as the $m$th absorption mode. From the practical point of view, absorbers with minimal coating layer thicknesses are of considerable interest. To this end, the mode $m = 0$ is most frequently used in the literature. In this case, the optimal layer thickness and its extinction coefficient are $d_l = \lambda/4n_l$ and $\kappa_l = 2/\pi \approx 0.64$, respectively.

Obviously, the PEC model has its limitations. It works quite well at long wavelengths, starting from the mid-IR range. In this case, the reflection coefficient of real metals at normal incidence of a wave is close to unity, and the phase of the reflected wave is close to $\pi$, as in PEC. In the near-IR range and especially in the visible range, the difference between these parameters for PEC and for real metals becomes significant if the interference phenomena are taken into account. This also affects the conditions of complete absorption for the Dallenbach layer. For instance, Kats et al. \cite{Kats2013} showed that for germanium coatings on a gold substrate, the optimal thickness in the visible range turns out to be several times less than $\lambda / 4n_l$. Park et al. \cite{Park2015OptLett,Park2014ACSPhot} presented a detailed theoretical analysis for the case of a substrate with optical constants that modulate real metals and for coating materials with $n_l > 1$. The authors of Ref. \cite{Park2015OptLett} came to the conclusion that in the case of real metals, the optimal extinction coefficient of the coating material, $\kappa_l$, exceeds the value of 0.64. These values of the complex refractive index are well matched by various semiconductor materials. Cleary et al. \cite{Cleary2015OMEx} have demonstrated strong IR absorption in coatings based on gallium-doped zinc oxide (GZO) on a silver substrate. GZO is an example of materials with a refractive index and extinction coefficient less than unity and therefore can be classified as an ENZ materias. Badsha et al. \cite{Badsha2014OptComm} analyzed in detail the conditions of total absorption for such coatings in the case of grazing incidence. However, the case of normal incidence, which is of greatest practical interest, has never been analyzed theoretically. 

In this paper, we consider absorption of electromagnetic radiation by a single-layer lossy coating on top of a metal substrate. Using numerical calculations, we find all possible combinations of the complex refractive index ($n_l+i\kappa_l$) of the coating material and its thickness ($d_l$), providing complete absorption of a plane monochromatic electromagnetic wave for the idealized case of the PEC substrate. Based on this analysis, we show that the condition of unity absorption described by Eq. (\ref{eq:Dbch}) is a special case corresponding to large values of $n_l$, i.e. $n_l \gg 1$. We also demonstrate that the unity absorption can be achieved for any values of $n_l$, including $n_l$ near zero. In the latter case, analytical expressions are also obtained for the conditions of total absorption: $d_l = \lambda/2 n_l$ and $\kappa_l = n_l^2/\pi$. An example of the design of light-absorbing structures based on real index-near-zero material is provided.

\section{Conditions for unity absorption}

Consider the interaction of a plane electromagnetic wave having a wavelength $\lambda$ with the model structure shown in Fig. \ref{fig_1}. The wave is incident on the structure at an angle $\theta_a$ measured from the normal. In the case of opaque metal substrates, absorptance can be calculated by the expression 
\begin{equation}
   A = 1 - R,
   \label{eq_2}
\end{equation}
where $R$ is the reflectance. In turn, reflectace can be calculated via the amplitude reflection coefficient $r$ as $R = |r|^2$. The expression for $r$ has the form 
\begin{equation}
    r = \frac{r_{al} + r_{ls} e^{2i \phi_l}}{1 + r_{al}r_{ls} e^{2i \phi_l}},
    \label{eq_3}
\end{equation}
where $r_{al}$ and $r_{ls}$ are the Fresnel reflection coefficients for the interfaces between the air and the coating layer and between the coating layer and the substrate, respectively; $\phi_l = 2\pi \widetilde{n}_l d_l \cos \theta_l/\lambda$; and $\theta_l$ is the complex angle which is determined through the angle of incidence $\theta_a$ via Snell's law $\widetilde{n}_a \sin \theta_a = \widetilde{n}_l \sin \theta_l$. We also define $\delta_l = n_l d_l/\lambda$ as a reduced coating thickness which will be used below in the analysis. 

\begin{figure}
\centering
\begin{tabular}{cc}
\includegraphics[width=0.4\linewidth]{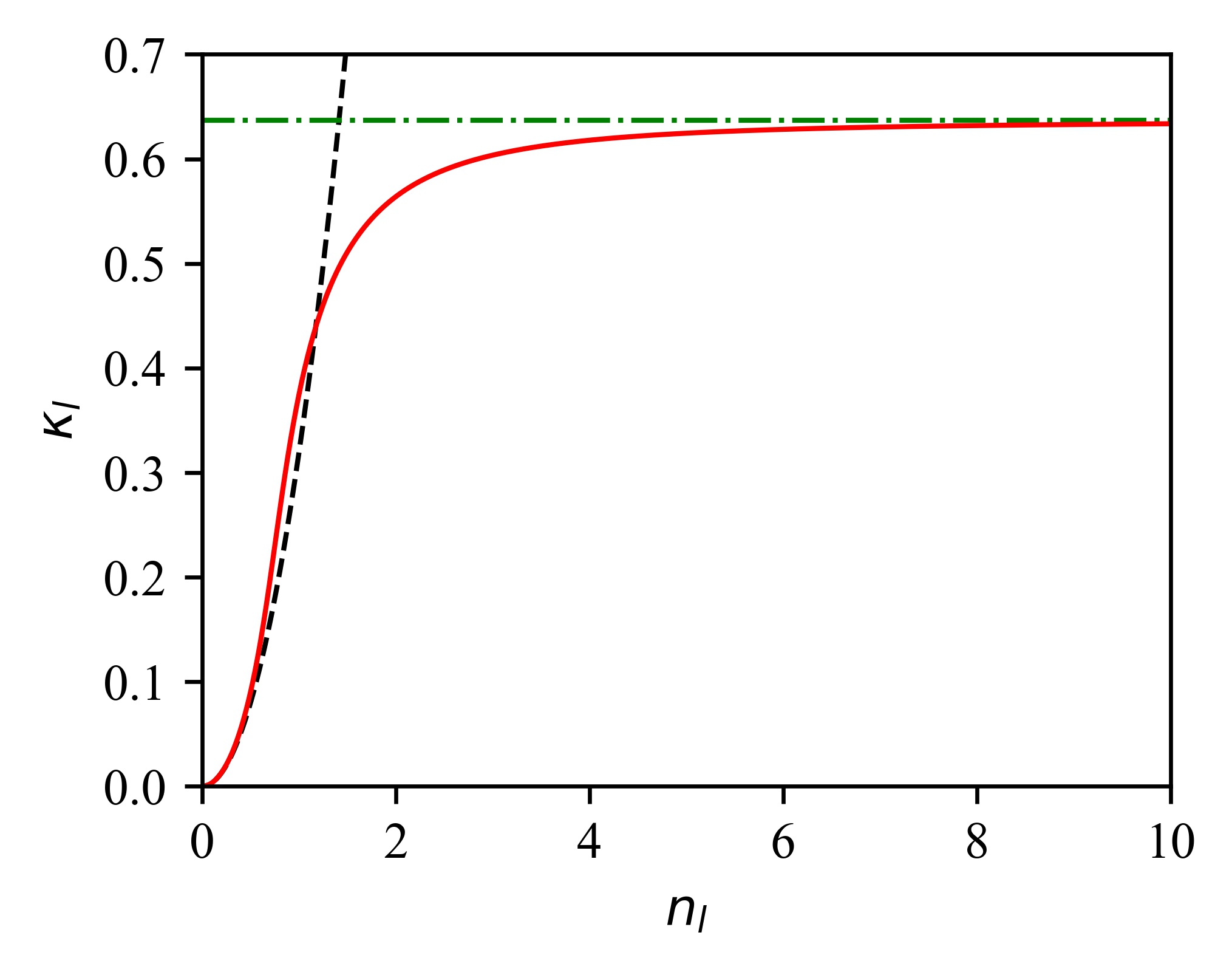} &
\includegraphics[width=0.4\linewidth]{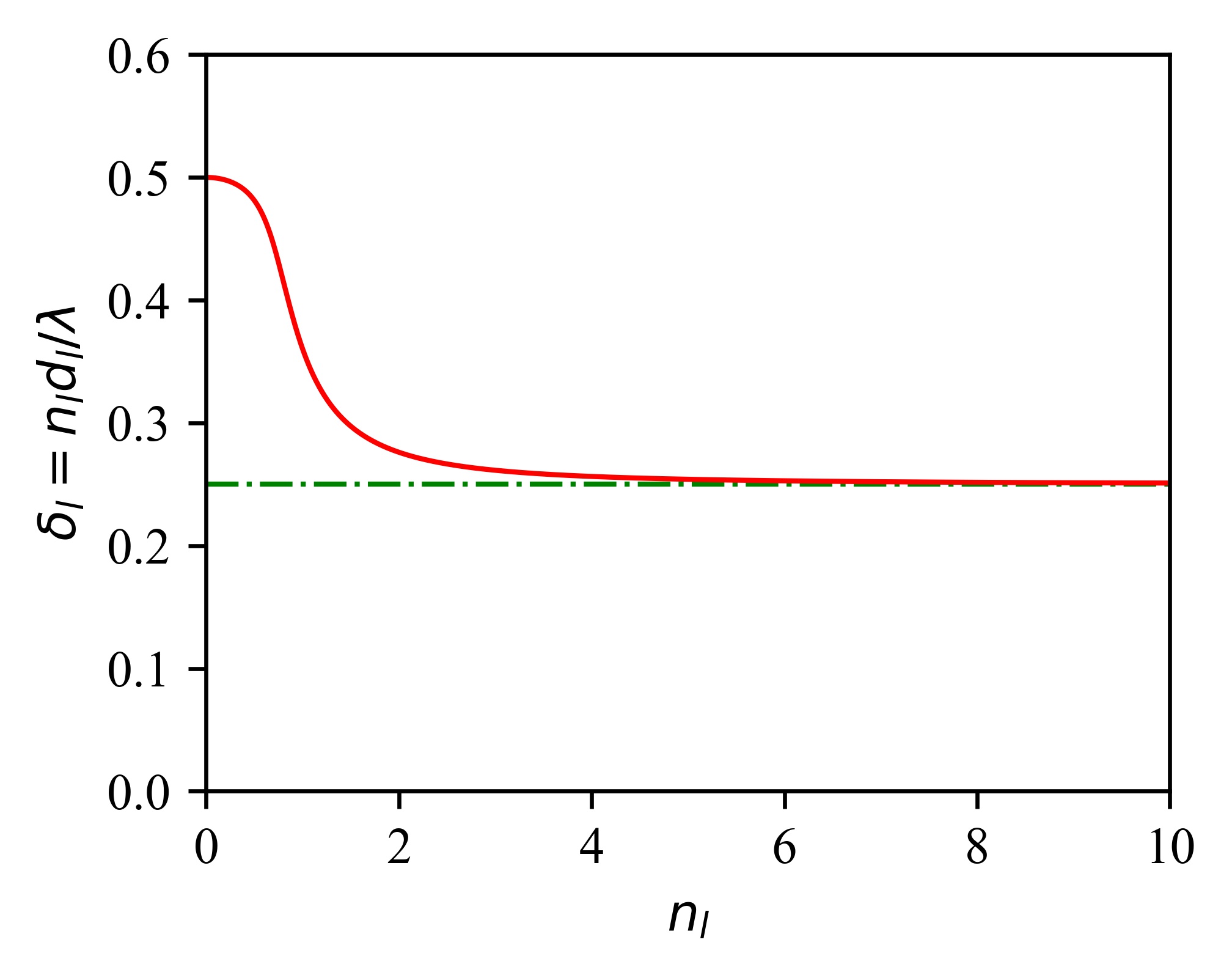} \\
(a) & (b)
\end{tabular}
\caption{(a)-(b): Solid red curves demonstrate the calculated combinations of parameters ($n_l, \kappa_l, \lambda_l$) of the Dallenbach layer on a PEC substrate, providing complete absorption of a normally incident plane wave. Horizontal green dot-dashed lines are classic solutions for the Dallenbach layer represented by Eqs. 1 at $m=0$. (a) Dependence of the optimal extinction coefficient $\kappa_l$ of the coating material on its refractive index $n_l$. Black dashed curve shows the fitting function $\kappa_l = 0.32 n_l^2$. (b) Dependence of the optimal normalized coating thickness $\delta_l$ on the refractive index of the coating material $n_l$. }
\label{fig_2}
\end{figure}

Let us analyze an idealized structure with a PEC substrate. In this case, $r_{ls} = -1$ in Eq. \ref{eq_3}. First, we consider the normal incidence of electromagnetic radiation, i.e. $\theta_a = 0$. The condition for unity absorption
is obtained from the roots of the numerator of Eq. \ref{eq_3}:
\begin{equation}
    r_{al} - e^{2i \phi_l} = 0,
    \label{eq_4}
\end{equation}
which can be rewritten in the form of two equations that must be simultaneously satisfied:
\begin{subequations}
\begin{eqnarray}
|r_{al}| = \exp(-4\pi \delta_l \kappa_l/n_l),
\end{eqnarray}
\begin{equation}
\arg(r_{al}) + 2\pi m = 4\pi \delta_l,  m = 0,1,2,...
\end{equation}
\label{eq_5}
\end{subequations}
Equations \ref{eq_5} can be solved numerically for $\delta_l$ and $\kappa_l$ at a fixed value of $n_l$. Thus, by varying $n_l$, one can find all possible combinations of ($n_l$, $\kappa_l$, $\delta_l$) corresponding to unity absorption. Figures \ref{fig_2}(a)-(b) show a solution space for $m = 0$. The green horizontal dot-dashed lines in Figs. \ref{fig_2}(a)-(b) denote the previously known solution defined by Eqs. \ref{eq_1}, i.e. $\kappa_l = 2/\pi$ and $\delta_l = 0.25$. It is seen that the numerical solution obtained here asymptotically tends to the indicated values only for large values of $n_l$. For $n_l < 2$, the numerical solution differs significantly from that defined by Eqs. \ref{eq_1}. As $n_l$ tends to zero, $\delta_l$ tends to a constant value of 0.5, and $\kappa_l$ tends to zero approximately as $0.32n_l^2$ (see black dashed curve in Fig. \ref{fig_2}(a)).  

\begin{figure}
\centering
\begin{tabular}{ccc}
\includegraphics[width=0.33\linewidth]{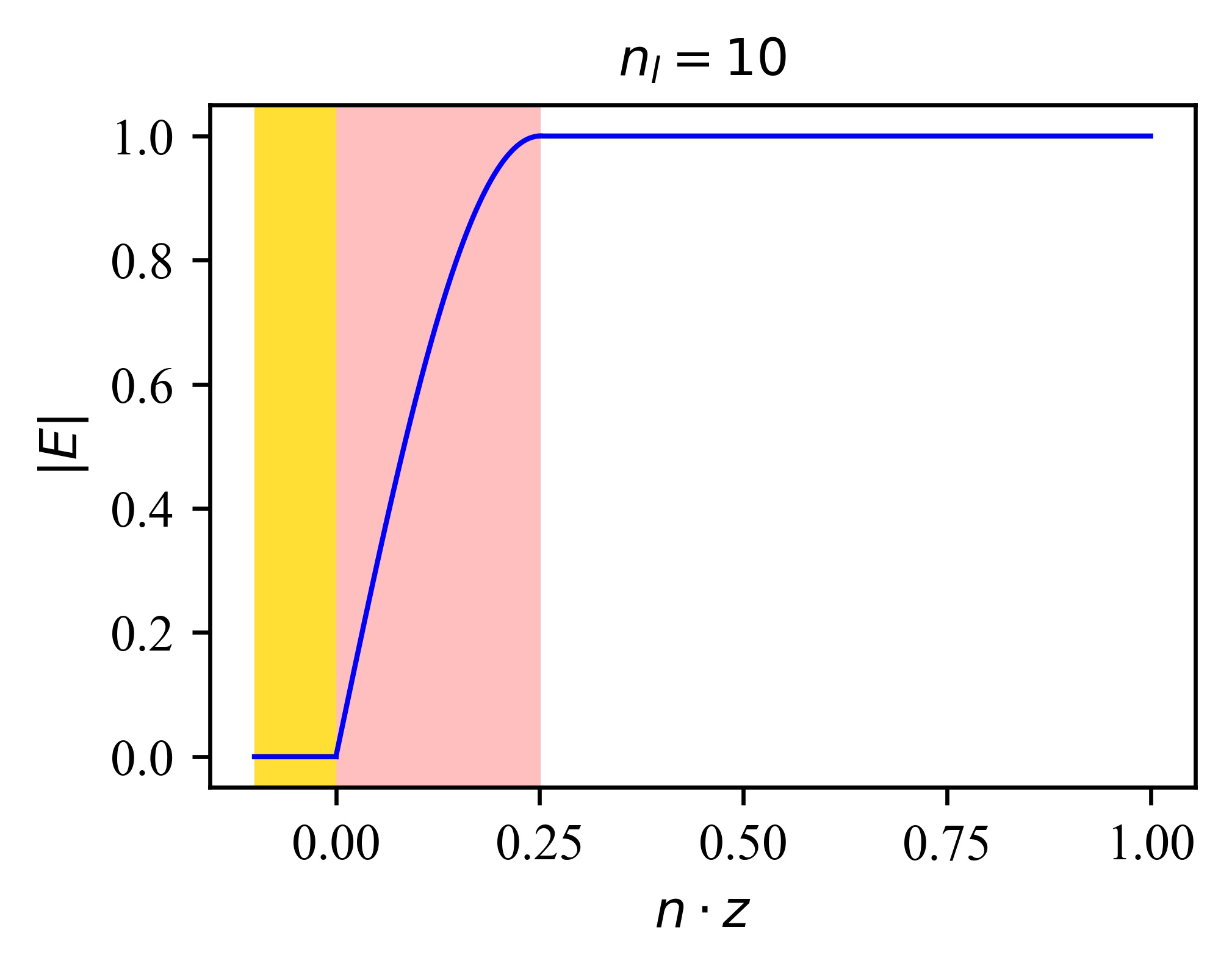} & \includegraphics[width=0.33\linewidth]{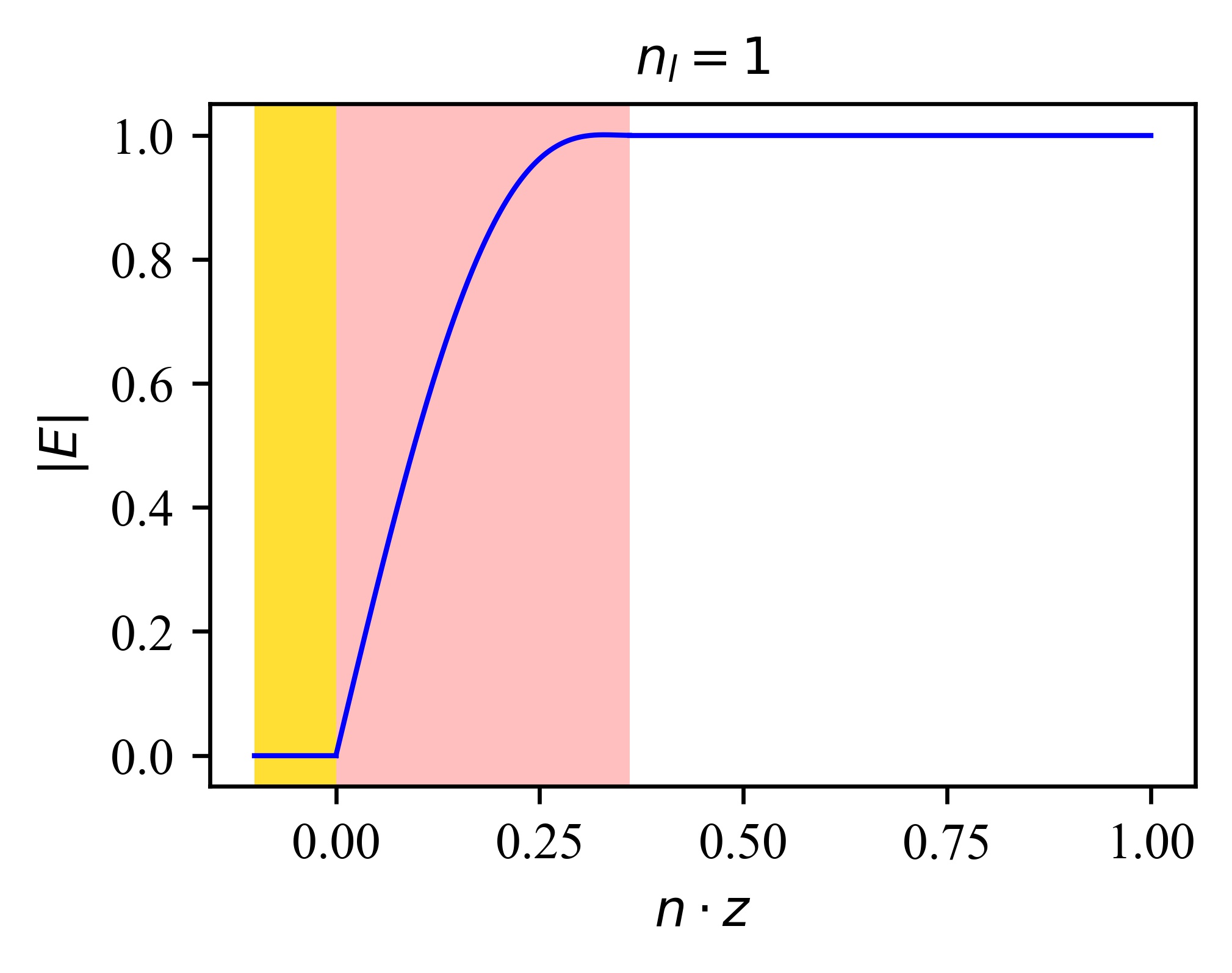} & \includegraphics[width=0.33\linewidth]{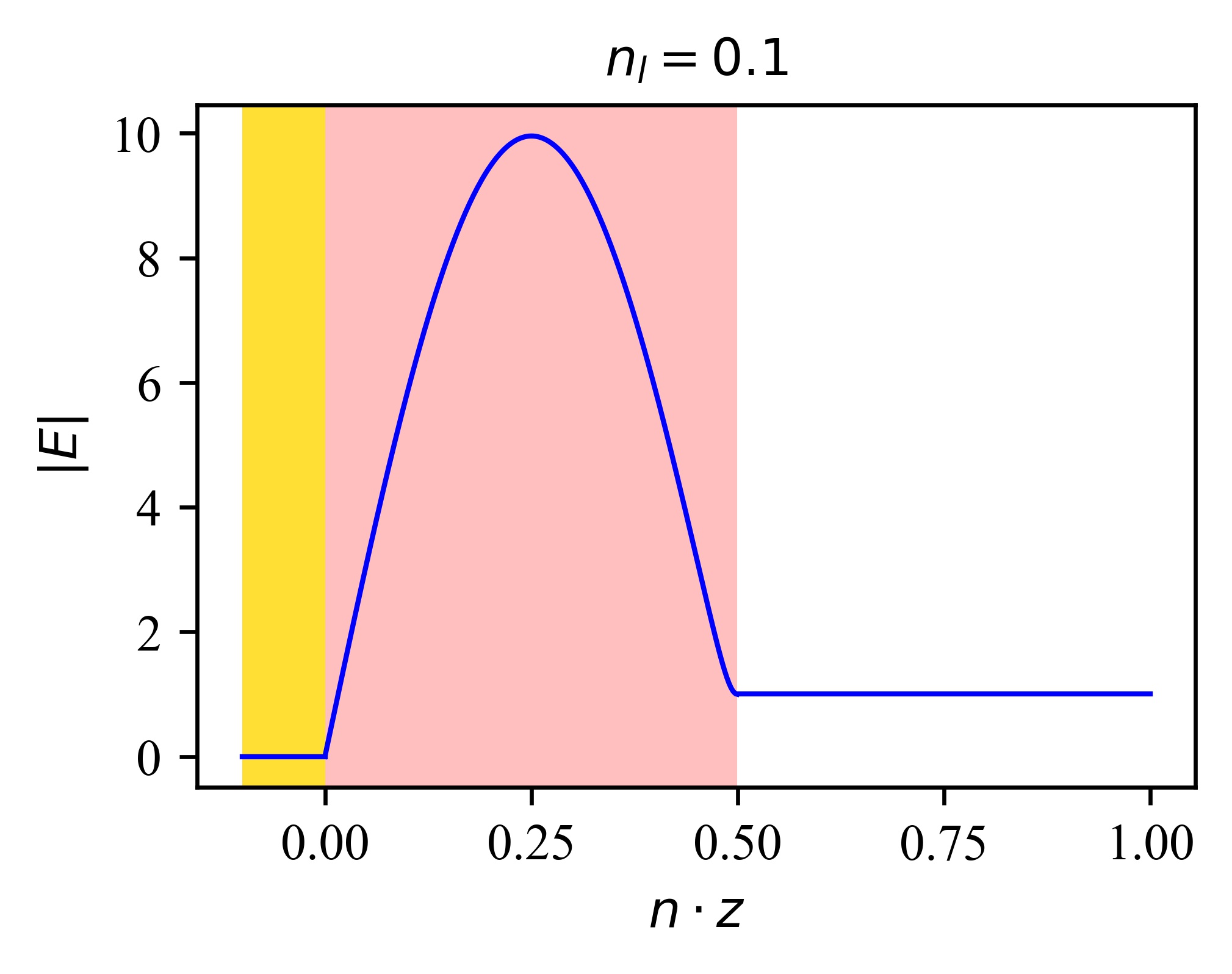}\\
(a) & (b) & (c) 
\end{tabular}
\caption{(a)-(c): Calculated profiles of the electric field for Dallenbach layers with different values of the refractive index of the coating material: (a) $n_l = 10$, (b) $n_l = 1$, and (c) $n_l = 0.1$.}
\label{fig_3}
\end{figure}

\begin{figure}
\centering
\begin{tabular}{ccc}
\includegraphics[width=0.33\linewidth]{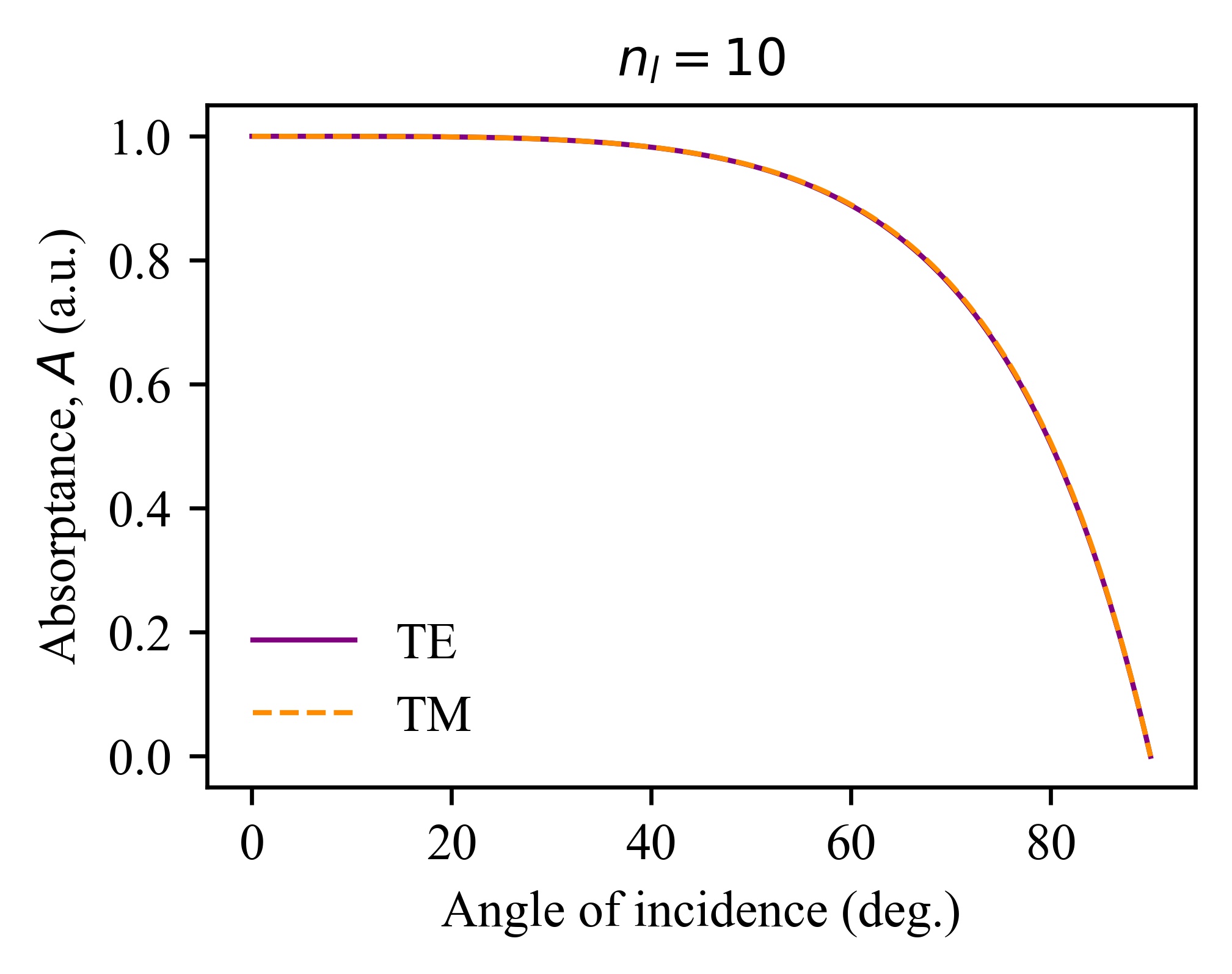} & \includegraphics[width=0.33\linewidth]{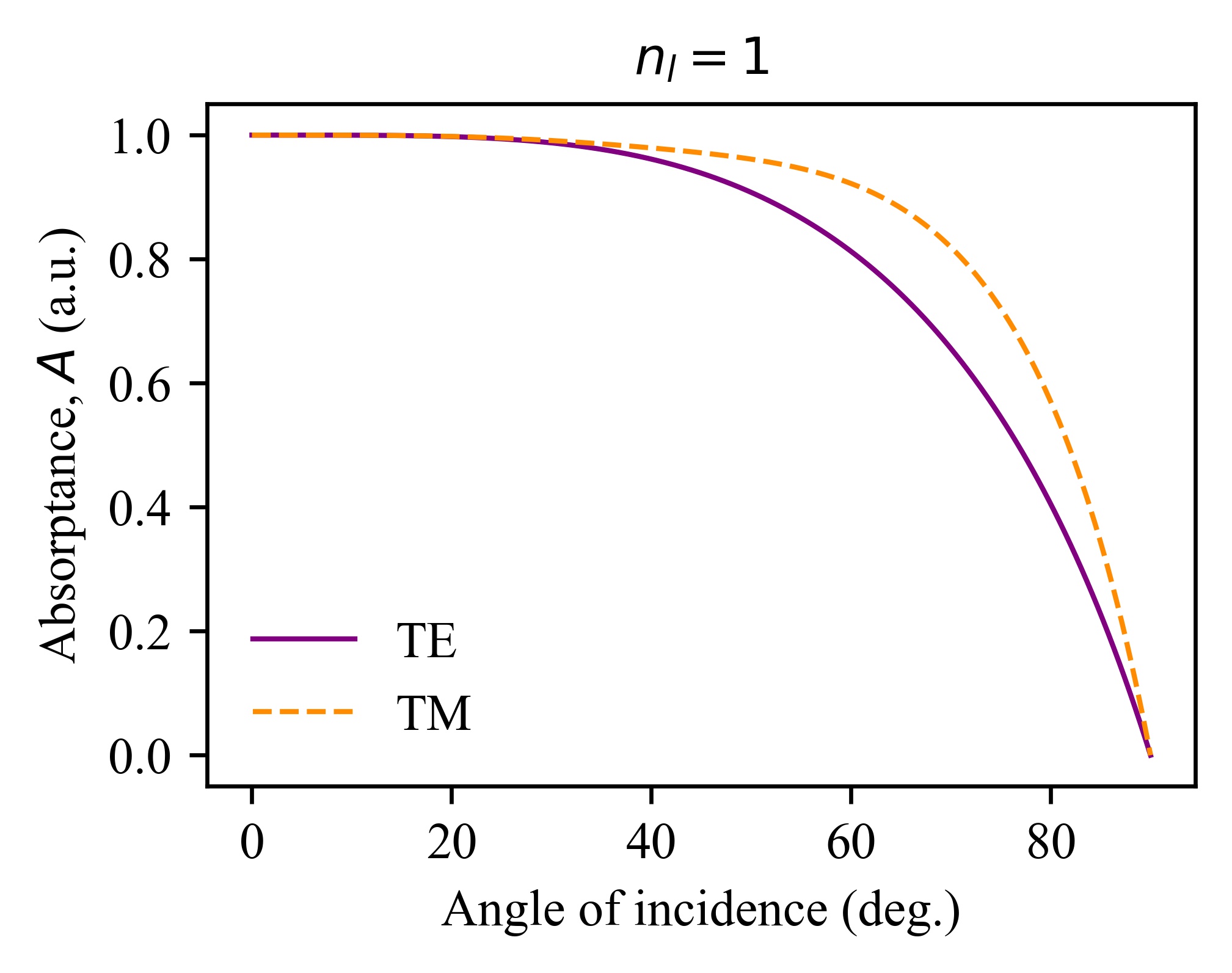} & \includegraphics[width=0.33\linewidth]{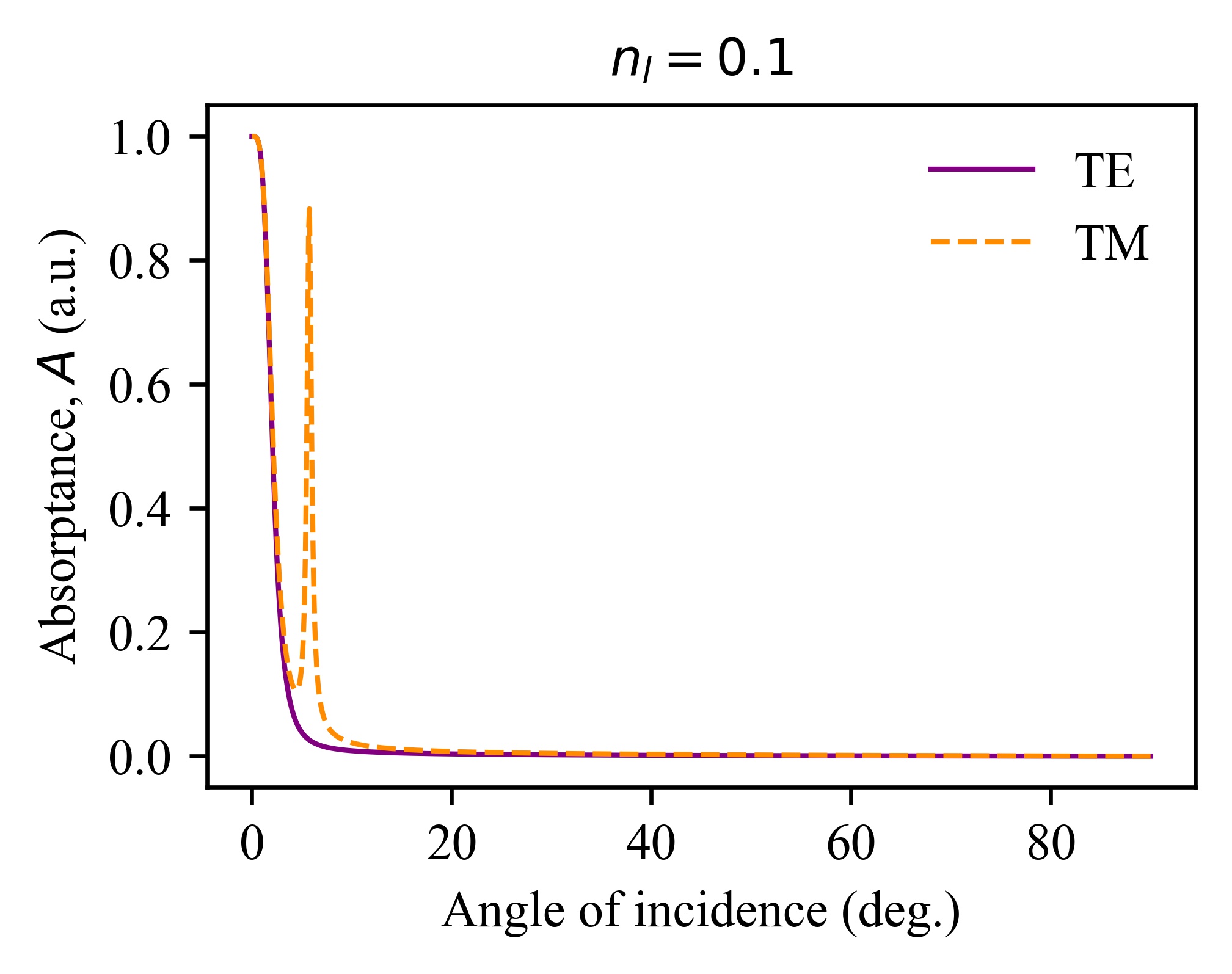}\\
(a) & (b) & (c)
\end{tabular}
\caption{(a)-(c): Calculated angular dependencies of the absorbance for Dallenbach layers with different values of the refractive index of the coating: (a) $n_l = 10$, (b) $n_l = 1$, and (c) $n_l = 0.1$. Dashed orange curves correspond to TM polarization, and solid purple curves correspond to TE polarization.}
\label{fig_4}
\end{figure}

Let us analyze Eq. \ref{eq_5}(a) under the conditions $n_l \rightarrow 0$, $\kappa_l < n_l$ and $\delta_l \rightarrow 0.5$. In this case, we obtain the following approximation for the Fresnel reflection coefficient for the upper interface of the structure: $r_{al} = (1 - \widetilde{n}_l)/(1 + \widetilde{n}_l) \approx (1 - n_l)/(1 + n_l) \approx 1 - 2n_l$. Equation 5a then transforms into $1 - 2n_l = \exp(-2\pi \kappa_l/n_l)$. Expanding the exponent in this equation into a Taylor series in powers of $\kappa_l/n_l$, we obtain the solution $\kappa_l \approx n_l^2/\pi \approx 0.32 n_l^2$.

It is also interesting to analyze the distribution of the electric field inside the considered absorber. The field within a layer can be calculated using the expressions: 
\begin{subequations}
\begin{eqnarray}
E_a(z) = E_0 e^{-i k_a z} + r E_0 e^{i k_a z}, z > d_l
\\
E_l(z) = \frac{2i t_{al} e^{i (k_l - k_a)d_{l}} E_0}{1 + r_{al}r_{lm} e^{2i k_{l}d_{l}}} \sin(k_l z), 0 < z \leq d_l
\\
E_s(z) = 0, z \leq 0,  
\end{eqnarray}
\label{eq_6}
\end{subequations}
where $ E_0 $ is the amplitude of the incident wave; $k_j = 2\pi \widetilde{n}_j/\lambda$; and $j = a,l,s$. Figures \ref{fig_3}(a)-(c) show the calculated profiles of the modulus of the electric field for three structures with different values of the refractive index: $n_l = 10$, $n_l = 1$, and $n_l = 0.1$. The values of $\delta_l$ and $\kappa_l$ were chosen in such a way that all three structures provide complete absorption of the incident wave. It follows from Fig. \ref{fig_3}(c) that in the case of $n_l < 1$, the field is enhanced inside the absorbing layer. Analysis of Eq. \ref{eq_6} at $n_l \rightarrow 0$, $\kappa_l \approx n_l^2/\pi$ and $\delta_ l\rightarrow 0.5 $ shows that the field amplitude inside the layer inversely depends on $n_l$. 

Let us now analyze the angular dependence of absorptance by the structures in question at different values of the refractive index. Figures \ref{fig_4}(a)-(c) show the calculated angular dependences of the absorption coefficient for three structures with $n_l = 10$, $n_l = 1$, and $n_l = 0.1$. The values of $\delta_l$ and $\kappa_l$ were chosen in such a way that all three structures provide complete absorption of the wave at normal incidence. The purple curves in Figs. \ref{fig_4}(a)-(c) correspond to the TE polarization of the incident plane wave, and orange ones, to the TM polarization. It follows from Figs. \ref{fig_4}(a)-(c) that for $ n_l = 10 $ and $ n_l = 1 $, absorption decreases relatively slow with increasing angle $ \theta_a $. When $ n_l = 0.1 $, strong absorption occurs only at incidence angles of several degrees, which is due to the fact that at $n_l < 1$, total external reflection at oblique incidence occurs at angles $\theta_a > \arcsin (n_l)$. In particular, for $n_l = 0.1$, total external reflection occurs for angles of incidence exceeding $\arcsin (0.1) \approx 5.74^\circ$. In the angular dependence of the absorptance for the TM wave, a narrow peak is observed in the region of the angle $ \theta_a \approx 5.73^\circ $, which is close in value to the Brewster angle for the upper structure interface, i.e. $\arctan (0.1) \approx 5.71^\circ$.

\section{Real materials}

\begin{figure}
\centering
\begin{tabular}{ccc}
\includegraphics[width=0.33\linewidth]{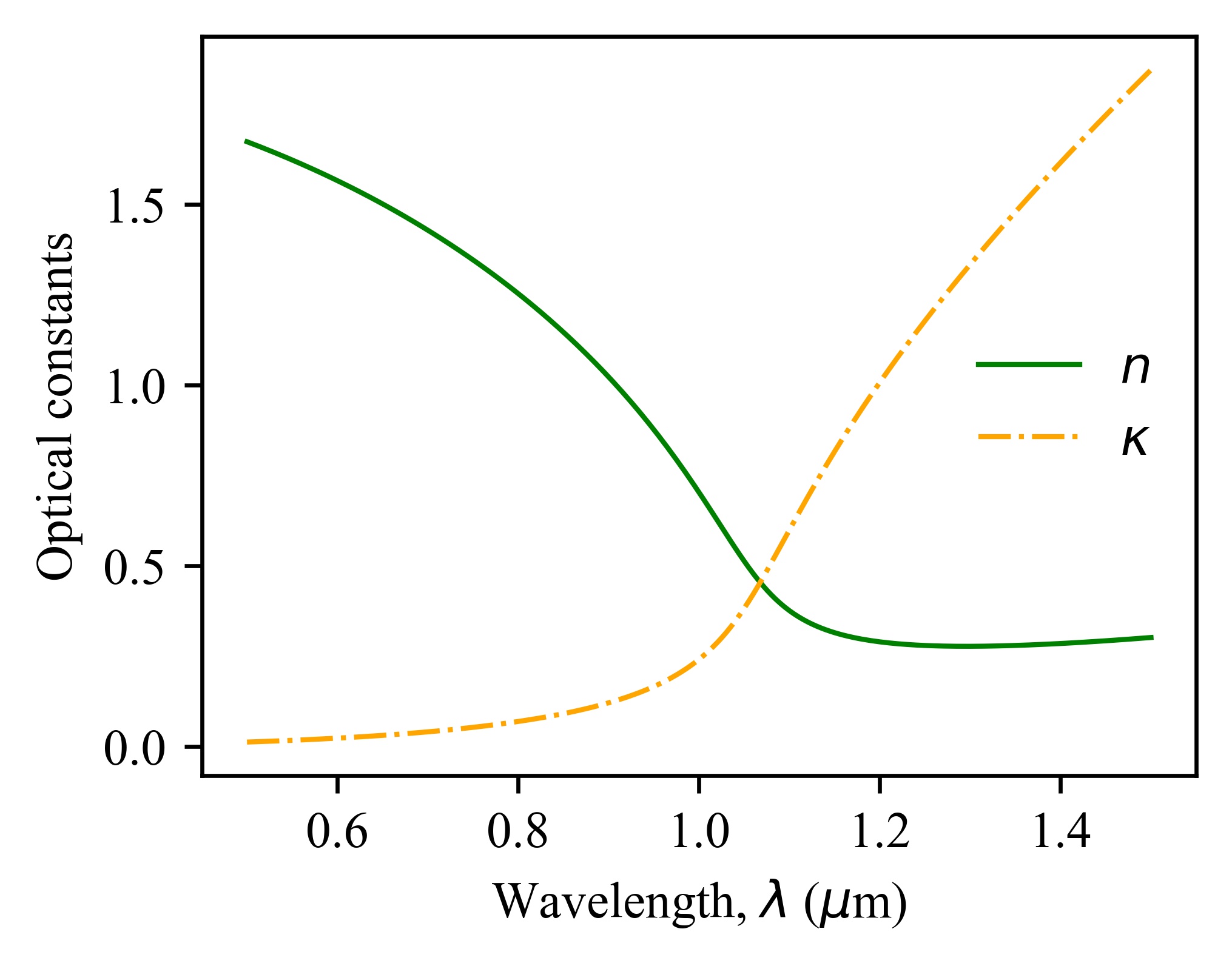} & \includegraphics[width=0.33\linewidth]{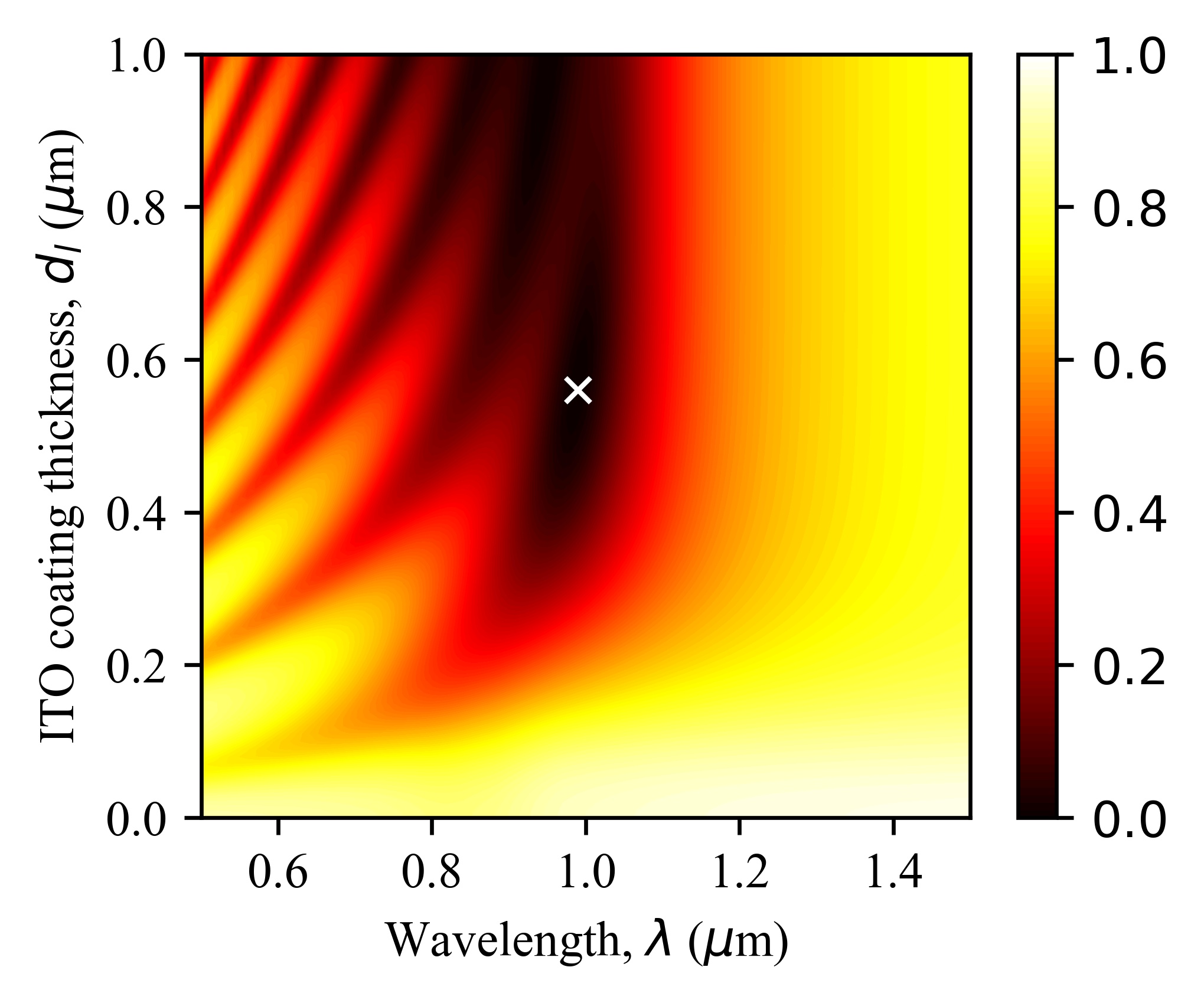} & \includegraphics[width=0.33\linewidth]{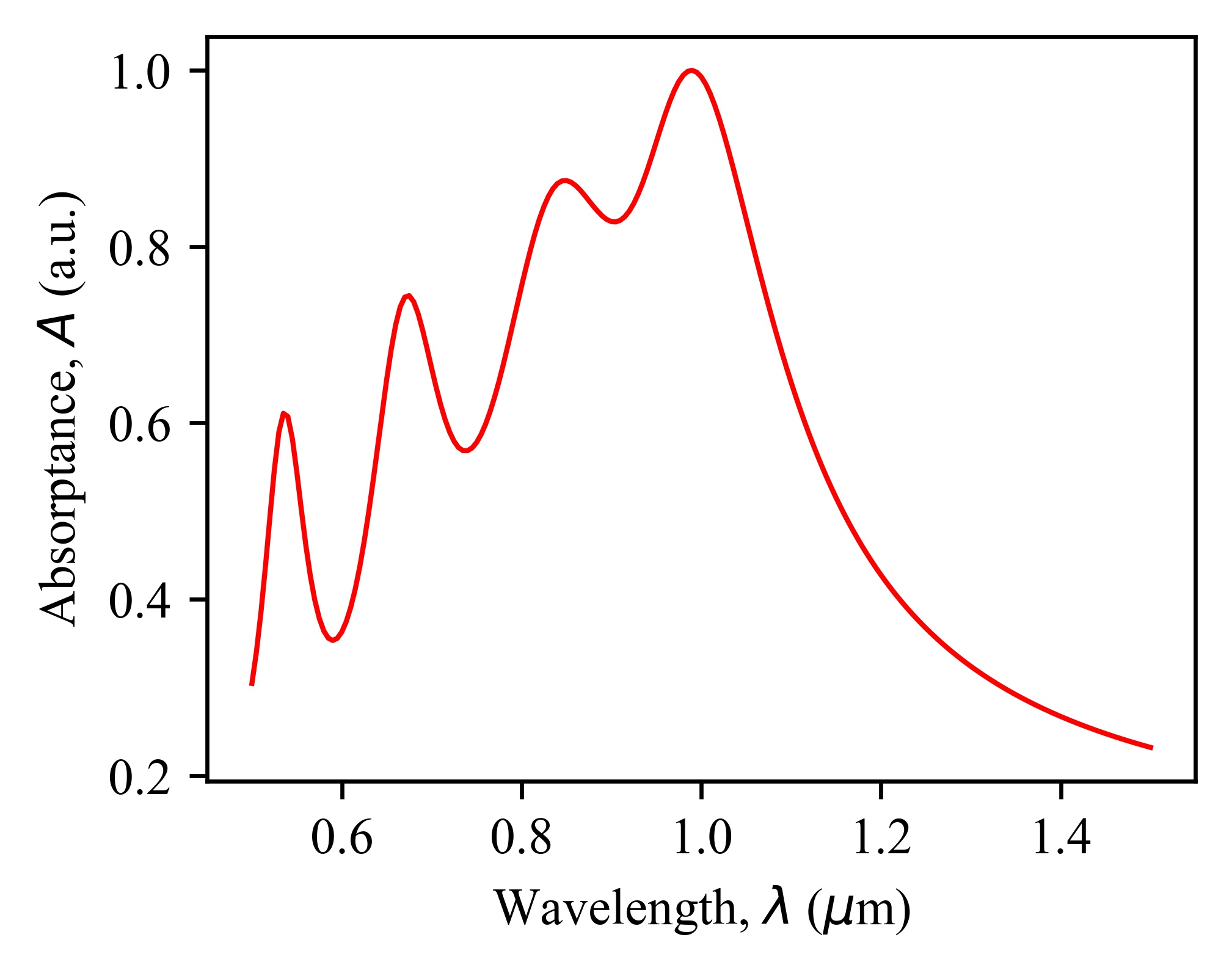}\\
(a) & (b) & (c) 
\end{tabular}
\caption{(a) Refractive index (solid green curve) and extinction coefficient (dot-dashed orange curve) of ITO at $N_e = 10^{21}$ cm$^{-3}$. (b) Reflectance of the ITO coating on the Al substrate as a function of the wavelength and thickness of the ITO ($N_e = 10^{21}$ cm$^{-3}$) coating. (c) Absorption spectrum of the ITO ($N_e = 10^{21}$ cm$^{-3}$) coating on the Al substrate for the coating having a thickness of 0.56 $\mu$m.}
\label{fig_5}
\end{figure}

The results presented in Fig. \ref{fig_1}(b) show that light-absorbing coatings can be made of a variety of materials. Coatings based on materials with refractive indices greater than unity have been extensively studied in previous works \cite{Kats2013, Park2014ACSPhot, Park2015OptLett, Wang2018Small, Liu2016NanoResearch, Wen2016SCM,Mirshafieyan2014OPEx, Mirshafieyan2016IEEE,Krayer2018ACSPhot, Dias2018AdvOptMat}. Here we consider in detail coatings based on ENZ materials. Examples of such materials are transparent conductive oxides such as indium tin oxide (ITO) and aluminum zinc oxide (AZO). Such oxides are usually characterized by high electron concentrations, $N_e \sim 10^{19} - 10^{21}$ cm $^{-3}$, resulting from the deposition conditions. Such high concentrations of charge carriers make it possible to obtain the ENZ mode in the infrared range. The optical properties of such materials are well described by the Drude model: 
\begin{equation}
    \varepsilon = \varepsilon_{\infty} - \frac{\omega_{p}^{2}}{\omega^2 + i \omega \Gamma},
    \label{eq_7}
\end{equation}
where $\varepsilon$ denotes the complex permittivity of the material at a frequency $\omega = 2\pi c/\lambda$, $\varepsilon_{\infty}$ is the permittivity at an infinite frequency, $\omega_p$ is the electron plasma frequency, and $\Gamma$ is the electron collision frequency. The last two quantities can be calculated by the formulas: $\omega_p = \sqrt{N_e e^2/m^* \varepsilon_0}$ and $\Gamma = e/m^* \mu$. Here $e$ is the elementary charge, $m^*$ is the electron effective mass, $\mu$ is the electron mobility, and $\varepsilon_0$ is the permittivity of free space.

Let us analyze the characteristics of ITO-based coatings. The parameters required for calculating the ITO optical constants by the Drude model were borrowed from Ref. \cite{Anopchenko2018ACSPhot}. We choose aluminum (Al) as the substrate material. The optical constants of Al for the calculations were borrowed from Ref. \cite{Rakic1995ApplOpt}. To begin with, let the concentration of electrons be $N_e = 10^{21}$ cm$^{-3}$. Figure \ref{fig_5}(a) shows the complex refractive index of ITO, calculated according to the Drude model at a given electron concentration for wavelengths in the range of 0.5 - 1.5 $\mu$m. One can see from Fig. \ref{fig_5}(a) that the ENZ mode is realized near $\lambda = 1$ $\mu$m. Figure \ref{fig_5}(b) demonstrates the calculated reflectance of the ITO/Al structure as a function of the wavelength and thickness of the ITO coating. According to the calculations, at point with a wavelength of $\lambda_0 \approx 0.99$ $\mu$m and a thickness of $d_0 \approx 0.56$ $\mu$m, the reflection vanishes, which is equivalent to total absorption. The complex refractive index of ITO at a given electron concentration at a wavelength of 0.99 $\mu$m is $\widetilde{n}_{\text{ITO}} = n_{\text{ITO}} + i \kappa_{\text{ITO}} = 0.74 + 0.22i$. According to the above predictions, the optimal extinction coefficient for a structure with a PEC substrate is $n_{\text{ITO}}^2/\pi \approx 0.18$, which is close to the extinction coefficient of the considered material $\kappa_{\text{ITO}} = 0.22$. The normalized optical thickness of the coating in this case is $\delta = d_0 n_{\text{ITO}}/\lambda_0 \approx 0.42$, which is also close to the analytical predictions for the case with the PEC substrate, i.e. $\delta = 0.5$. Note that there are also other points with a larger coating thickness and lower wavelength values, at which zero reflection is implemented. However, we restrict ourselves to considering structures with a minimum resonant coating thickness. The calculated absorption spectrum for a structure with an ITO coating having a thickness of 0.56 $\mu$m is shown in Fig. \ref{fig_5}(c). 

\begin{figure}
\centering
\begin{tabular}{cc}
\includegraphics[width=0.33\linewidth]{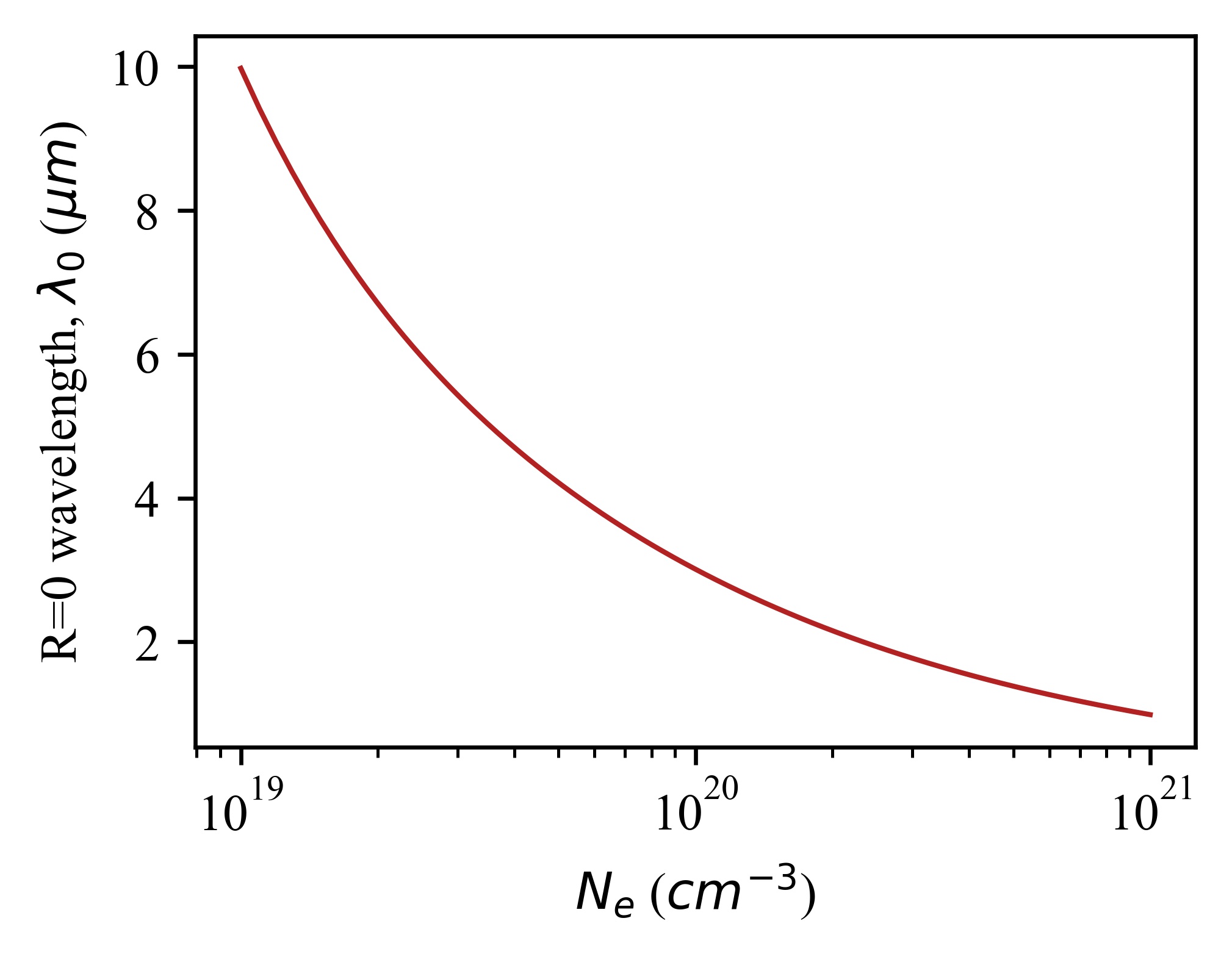} & \includegraphics[width=0.33\linewidth]{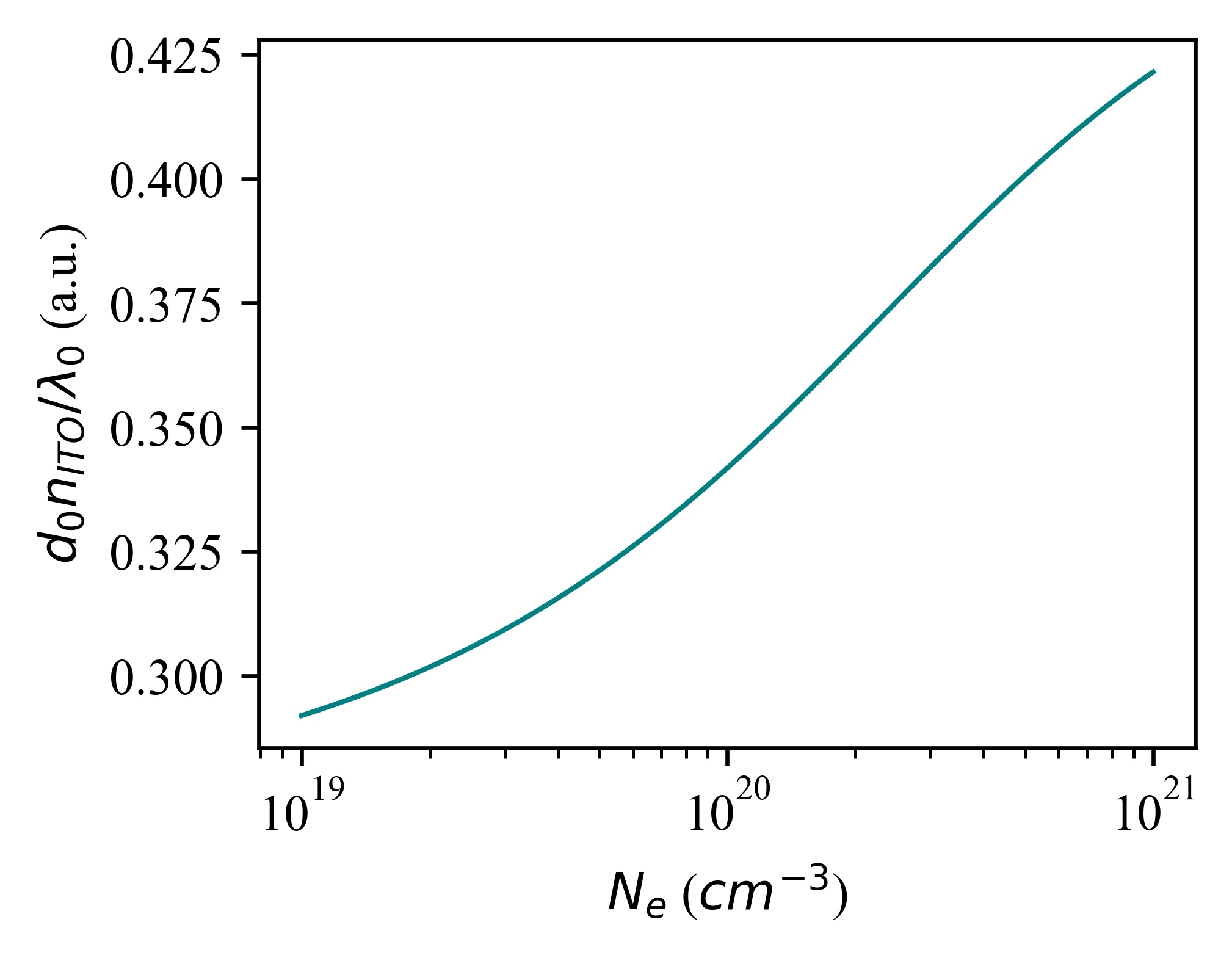}\\
(a) & (b) 
\end{tabular}
\caption{(a) Dependence of the resonant wavelength $\lambda_0$ for the ITO coating atop the Al substrate on the electron concentration in ITO. (b) Dependence of the normalized optical thickness $d_0 n_{\text{ITO}}/\lambda_0$ of the ITO coating atop the Al substrate on the electron concentration
 in ITO.}
\label{fig_6}
\end{figure}

What happens when the concentration of electrons in the coating material changes? For each value of Ne, we can also find a point at which the reflection vanishes. Figure \ref{fig_6} shows the dependences of $\lambda_0$ and $d_0 n_{ITO}/\lambda_0$ on $N_e$ in the range $10^{19} - 10^{21}$ cm$^{- 3}$. It can be seen from Fig. 6 that with a decrease in the electron concentration from $10^{21}$ cm$^{-3}$ to $10^{19}$ cm$^{-3}$, the resonance wavelength increases from approximately 1 $\mu$m to 10 $\mu$m, and the normalized optical thickness of the coating will decrease from approximately 0.42 to 0.29. Note also that changes in the substrate material do not lead to significant quantitative changes in the values of $\lambda_0$ and $d_0$. 

\section{Conclusions}

We have investigated absorption of electromagnetic radiation in the so-called Dallenbach layer, resulting from the phenomenon of destructive interference. We have shown that the Dallenbach layer can provide complete absorption of a normally incident plane wave in a wide range of complex refractive indices of the coating layer material, including close-to-zero indices. This fact makes it possible to design efficient ENZ-based absorbers. Simple analytical design rules have been derived, demonstrating that such structures provide high angular absorption selectivity, which is determined by total external reflection. An example of the design of a tunable wavelength absorber based on an ITO layer on an Al substrate have been presented.

\bibliography{aipsamp}

\end{document}